\documentstyle[12pt,epsfig]{article}

\def\sbotL{\tilde{b}_L}
\def\sbotR{\tilde{b}_R}
\def\stopL{\tilde{t}_L}

\def\sbota{\tilde{b}_1}

\def\stopa{\tilde{t}_1}

\def\msbota{m_{\tilde{b}_1}}
\def\msbotb{m_{\tilde{b}_2}}
\def\mstopa{m_{\tilde{t}_1}}
\def\mstopb{m_{\tilde{t}_2}}
\def\sb{\sin^2\theta_b}
\def\cb{\cos^2\theta_b}
\def\st{\sin^2\theta_t}
\def\ct{\cos^2\theta_t}
\def\nonm{\nonumber}

\def \cp89{{\it CP Violation,} edited by C. Jarlskog (World Scientific,
Singapore, 1989)}

\def \f79{{\it Proceedings of the 1979 International Symposium on Lepton and
Photon Interactions at High Energies,} Fermilab, August 23-29, 1979,
ed. by T. B. W. Kirk and H. D. I. Abarbanel (Fermi National
Accelerator Laboratory, Batavia, IL, 1979}
\def \hb87{{\it Proceeding of the 1987 International Symposium on Lepton and
Photon Interactions at High Energies,} Hamburg, 1987, ed. by W. Bartel
and R. R\"uckl (Nucl. Phys. B, Proc. Suppl., vol. 3) (North-Holland,
Amsterdam, 1988)}

\def \ichep72{{\it Proceedings of the XVI International Conference on High
Energy Physics}, Chicago and Batavia, Illinois, Sept. 6 -- 13, 1972,
edited by J. D. Jackson, A. Roberts, and R. Donaldson (Fermilab,
Batavia, IL, 1972)}

\def \ite{{\it et al.}}
\def \lkl87{{\it Selected Topics in Electroweak Interactions} (Proceedings of
the Second Lake Louise Institute on New Frontiers in Particle Physics,
15 -- 21 February, 1987), edited by J. M. Cameron \ite~(World
Scientific, Singapore, 1987)}
\def \ky85{{\it Proceedings of the International Symposium on Lepton and
Photon Interactions at High Energy,} Kyoto, Aug.~19-24, 1985, edited
by M.  Konuma and K. Takahashi (Kyoto Univ., Kyoto, 1985)}

\def \np#1#2#3{Nucl. Phys. {\bf#1}, #2 (#3)}

\def \pl#1#2#3{Phys. Lett. {\bf#1}, #2 (#3)}

\def \pr#1#2#3{Phys. Rev. {\bf#1}, #2 (#3)}

\def \prl#1#2#3{Phys. Rev. Lett. {\bf#1}, #2 (#3)}
\def \prp#1#2#3{Phys. Rep. {\bf#1}, #2 (#3)}

\def \zpc#1#2#3{Z. Phys. C. {\bf#1}, #2 (#3)}

\def \si90{25th International Conference on High Energy Physics, Singapore,
Aug. 2-8, 1990}
\def \slc87{{\it Proceedings of the Salt Lake City Meeting} (Division of
Particles and Fields, American Physical Society, Salt Lake City, Utah,
1987), ed. by C. DeTar and J. S. Ball (World Scientific, Singapore,
1987)}
\def \slac89{{\it Proceedings of the XIVth International Symposium on
Lepton and Photon Interactions,} Stanford, California, 1989, edited by
M.  Riordan (World Scientific, Singapore, 1990)}
\def \smass82{{\it Proceedings of the 1982 DPF Summer Study on Elementary
Particle Physics and Future Facilities}, Snowmass, Colorado, edited by
R.  Donaldson, R. Gustafson, and F. Paige (World Scientific,
Singapore, 1982)}
\def \smass90{{\it Research Directions for the Decade} (Proceedings of the
1990 Summer Study on High Energy Physics, June 25--July 13, Snowmass,
Colorado), edited by E. L. Berger (World Scientific, Singapore, 1992)}
\def \tasi90{{\it Testing the Standard Model} (Proceedings of the 1990
Theoretical Advanced Study Institute in Elementary Particle Physics,
Boulder, Colorado, 3--27 June, 1990), edited by M. Cveti\v{c} and P.
Langacker (World Scientific, Singapore, 1991)}

\def \zpc#1#2#3{Zeit. Phys. C {\bf#1}, #2 (#3)}

\begin{document}

\begin{titlepage}
\begin{flushright}
UCLA/96/TEP/2\\
hep-ph/9601392 \\
January 31, 1996\\
\end{flushright}

\vskip 2.cm

\begin{center}
{\Large\bf Speculations on ALEPH's Dijet Enhancement} 
\vskip 2.cm

{\large A. K. Grant, R. D. Peccei, T. Veletto and K. Wang}

\vskip 0.5cm

{\it  Department of Physics, University of California at Los
Angeles,\\Los Angeles, CA 90095-1547}
\vskip 3cm
\end{center}

\begin{abstract}
We interpret the dijet enhancement reported by the ALEPH collaboration in
the process $e^+e^-\rightarrow $4 jets as being due to the production
of a pair of bottom squarks, followed by their R-parity violating
decays into pairs of light quarks. Constraints on this speculative
interpretation are examined. Some of the consequences of our
hypothesis are drawn within the context of softly broken
supersymmetry.
\end{abstract}
\vfill
\end{titlepage}

Recently the ALEPH
collaboration presented a preliminary analysis of  about $6~\rm{pb}^{-1}$
of data collected at $\sqrt s= 130-136$ GeV at LEP \cite{ALEPH}.
Although cross sections for standard processes appear to be consistent
with expectations, they reported some peculiar 4-jet events in their
data for which there is no canonical interpretation. What ALEPH sees
is an excess of 4-jet events (14 observed, 7.1 expected), with 8 of
these events clustered at a  dijet ``sum mass'' of about 110 GeV. This
dijet sum mass is arrived at by combining together the 2 dijet masses in
the events which have the smallest mass difference between them.
Assuming that what is seen is a signal for something, it corresponds
to a cross section of $2.5 \pm 1$~pb. This ``signal'' is quite distinct
from what is expected from QCD 4-jet events, where the event
distribution is essentially flat in the dijet sum mass. Indeed, in the
sum mass bin from 102-116 GeV where the 8 events are clustered, one
expects only 1.35 events from QCD.

ALEPH makes no claims about this
signal and it could well be a statistical fluctuation which will
disappear as more data is collected. Nevertheless, since the dijet mass
difference in the analysis is restricted to be below 20 GeV, it is
tempting to speculate that the ALEPH excess is due to the production of
a pair of particles of mass of about 55 GeV which then each decay into
pairs of jets. In fact, already Farrar \cite{Farrar} has suggested
that the ALEPH events are associated with the pair production of
squarks, which subsequently decay into a quark and a light gluino.
Although the light gluino scenario is interesting, 
here we wish to speculate in another
direction. We also want to associate the ALEPH signal with the pair
production of squarks. However, in contrast to Farrar, we suggest that
what is produced is only the $\tilde{b}_L$ and that the dijets result
from the R-parity violating decay of this squark to pairs of light
quarks. In what follows we will try to justify this speculation and
draw some of its consequences.

For a 55 GeV squark at LEP 1.5 energies,
the cross section for $e^+e^-\rightarrow \tilde{b}_L\tilde{b}^*_L$ is
about 1 pb. Thus the size of the signal is of the right magnitude.
Although the electroweak symmetry does not permit an R-parity violating
trilinear term involving the $(t,b)_L$ doublet in the superpotential,
there is a  possible $\Delta B=1$  R-parity violating term involving $ b_R$
\cite{Zwirner}. Since the $\tilde{b}_L$ squarks mix with the
$\tilde{b}_R$ squarks, these latter couplings (if they are present) 
allow for the decay of $\tilde{b}_L$ into light quarks. One must
check, however, that this decay does not run afoul of any of the
existing constraints on the trilinear R-parity violating couplings.

There are nine possible $\Delta B=1$ trilinear R-parity violating couplings
in the superpotential involving right-handed isosinglet
quarks
\begin{equation}
\label{operator}
W=\lambda_{ijk}D^i_RD^j_RU^k_R,
\end{equation}
since $ \lambda_{ijk}= -\lambda_{jik}$. The most stringent
bound on these couplings is that on $\lambda_{dsu}$, which comes
from the process $N N \rightarrow K K X$, where $N$ is a nucleon
and $K$ is a particle with unit strangeness.  In Ref.~\cite{Goity},
it was shown that one could have a bound as strong
as $|\lambda_{dsu}|~<~10^{-7}$, but with large hadronic uncertainties and
significant model dependence.
Six of the couplings $\lambda_{ijk}$ will
involve a $b_R$:  $\lambda_{bdu}; \lambda_{bsu}; \lambda_{bdc}; 
\lambda_{bsc}; \lambda_{bdt}; \lambda_{bst}$. The last two couplings
are not relevant here since the top is too heavy to be produced
in the decay of a 55 GeV $\sbotL$.
However, assuming that $\tilde{b}_L$ is  55 GeV, one must make sure
that
\[
\lambda_{bdt}; \lambda_{bst}<< e
\]
to preserve the decay
$t\rightarrow W b$ as the main top decay mode.  To our knowledge, there
are no stringent bounds on $\lambda_{bsu}$, $\lambda_{bdc}$ 
and $\lambda_{bsc}$. 
However, $\lambda_{bdu}$  gives a contribution to $n- \bar{n}$
oscillations and the experimental limit on the neutron oscillation
lifetime \cite{PDG} of $\tau~>1.2~\times~10^8 $ sec bounds this
parameter to be below about $10^{-5}$ \footnote{The actual bound
depends in detail on the supersymmetric spectrum, as discussed by
Goity and Sher \cite{Goity}.}. As long as some of the couplings
$\lambda_{bij}$ are of order $ 10^{-5}$ or larger, the decay of the 
produced $\tilde{b}_L$
into light quarks will not produce a displaced vertex 
\footnote{ In fact, the decay of the $\tilde{b}_L$ via the $R$-parity
violating operator (\ref{operator}) can only occur in the  presence
of some (typically small) $\tilde{b}_L$-$\tilde{b}_R$ mixing.  Hence
we need $\lambda_{bij} \sin\theta_b \geq 10^{-5}$, where 
$\sin\theta_b\sim 5\times 10^{-2}$ is the sine of the 
$\tilde {b}_L$-$\tilde{b}_R$ mixing angle.  However, the combination 
$\lambda_{bij} \sin\theta_b$  also cannot be too large, for this would
lead to  $Z\rightarrow q q' \sbotL$ decays
at an unacceptable level.}.
Thus, the existence
of at least one R-parity violating
coupling of sufficient 
strength among $\lambda_{bsu}, \lambda_{bdc}$ and
$\lambda_{bsc}$ makes the suggested scenario
phenomenologically viable. It could be argued that we have introduced
unnatural hierarchies in the couplings $\lambda_{ijk}$ by requiring
that we evade certain bounds while still retaining one coupling of sufficient
strength.  However, it is worth pointing out that even within the minimal
standard model one has sizable hierarchies in the Yukawa sector, such as
$m_u/m_t\sim 10^{-5}$.  We should note that, at this stage, one could have
also imagined that the ALEPH events came from the pair production of the 
lightest stop, followed by their R-parity violating decays into light quarks. 
For this to be viable, however, one would have to imagine that the coupling 
$\lambda_{sdt}$ dominated over both $\lambda_{bst}$ and $\lambda_{bdt}$ to 
account for the lack of decays with a $b$-jet in the final state in the 
ALEPH 4 jet sample~\cite{ALEPH}. 

To proceed, however, we must still check whether our
suggestion is theoretically sound. For the decay 
$ \tilde{b}_L\rightarrow $ 2-jets to dominate, it is 
important that the $\tilde{b}_L$ be the LSP. In
particular, all neutralinos should be heavier than the $\tilde{b}_L$. Otherwise
the decay $\tilde{b}_L\rightarrow b  \chi_1^0$  is likely to be dominant, since
the relevant coupling is of $O(e)$. We have associated the ALEPH 
enhancement to the production of a $\tilde{b}_L$ because this squark, 
along with the stops,  has a mass that is sensitive to the large top 
Yukawa coupling. Thus, as is well known \cite{radiative}, even starting 
with universal SUSY-breaking scalar masses at  the GUT scale, 
the $\tilde{b}_L$ has the possibility of obtaining  a nonuniversal mass. 
In fact, assuming universal soft breaking masses for the scalars and the 
gauginos  and $\tan \beta\sim 2$, one finds
\begin{equation}
\label{universal_1}
m^2_{\tilde{d}_L} \simeq m_0^2 + 6.8 m^2_{1/2},
\end{equation}
while
\begin{equation}
\label{universal_2}
m^2_{\tilde{b}_L } \simeq 0.51 m_0^2 + 5.5 m^2_{1/2}.
\end{equation}
One sees from the above that it is possible to have the  
$\tilde{b}_L$  mass be smaller than that of the other down squarks, 
provided the universal scalar mass $m_0$ dominates 
over $m_{1/2}$. However, because we want $m_{\tilde{b}_L } \simeq 55$ GeV 
then both $m_0$ and, particularly, the universal gaugino mass $m_{1/2}$ 
must be quite small. This, in general, is unacceptable since it leads 
to one or more very light neutralinos in the spectrum. Neutralinos
in this mass range are 
excluded experimentally by LEP \cite{Nbounds}.  But more importantly,
the presence of neutralinos lighter than the $\sbotL$ would
alter the decays of these squarks in an undesirable way.
Indeed, in the presence of such light neutralinos, the weak decay
$\tilde{b}_L\rightarrow \chi^0_1 b$ followed by the $R$-parity violating
decay $\chi^0_1 \rightarrow 3$ jets would give a different
experimental signature from that suggested by the ALEPH data.

The presence of light neutralinos, however, is a feature of 
the particular pattern of SUSY  
breaking which one assumes at the GUT scale. 
The principal contributor to the $m_{1/2}$ piece of
$m_{\tilde{b}_L }$ in Eq.~(\ref{universal_2}) is the gluino component. 
However, the neutralino masses are sensitively dependent on the
SUSY breaking masses one gives to the {\bf electroweak gauginos}, but 
weakly  dependent on the gluino mass. If the electroweak  
gaugino masses are taken to be different from the gluino mass, 
it is possible to make the neutralinos sufficiently heavy by having 
the $ SU(2)\times U(1)$  gaugino masses $m_{1}$  and $m_2$ heavy. 
This, per se, should not affect terribly the $\tilde{b}_L $ mass. 
However, to be sure one must examine anew the spectrum of 
supersymmetric excitations one gets in the case where the 
soft breaking of supersymmetry involves nonuniversal gaugino masses, 
$m_{1} \neq m_{2} \neq m_3$
\footnote{ Nonuniversal gaugino masses have been advocated
recently by Roszkowski and Shifman\cite{RS} in a different context.}.

We present below the results of a study of a minimal supergravity model with  
unequal soft supersymmetry breaking mass parameters in the gaugino sector, 
but  where one still has a common SUSY breaking mass for all the scalars. 
To constrain the model further, we also assume that the electroweak 
symmetry is broken radiatively \cite{radiative}. Even though a 
common mass is assumed 
for all the scalars at the GUT scale, these masses evolve to 
different values at low energy as a result of radiative effects. 
For the squarks of the first two
generations and for the sleptons, the effective masses at low energy 
are sensitive functions of the gaugino masses and the evolution 
of the coupling constants.  The masses for the third generation 
squarks depend, in addition, on the top Yukawa coupling and 
its evolution. Furthermore,  for the mass squared of 
the $\tilde{t}_L$ and $\tilde{t}_R$ one cannot neglect the 
SUSY-preserving contribution of $m_t^2$, due to the large top mass.

Solving the renormalization group equations for the squark and 
slepton masses \cite{radiative}, one obtains the following 
approximate formulas for the light squarks and the sleptons: 
\begin{eqnarray*}
m^2_{\tilde{e}_L}(t) & = & m^2_0 -0.27\cos 2\beta M^2_Z 
+x_1(t)m^2_1 +x_2(t)m^2_2 \phantom{\frac{1}{2}}\\ 
m^2_{\tilde{\nu}_L}(t) & = & m^2_0 + 0.5\cos 2\beta M^2_Z 
+x_1(t)m^2_1 +x_2(t)m^2_2 \phantom{\frac{1}{2}}\\
m^2_{\tilde{e}_R}(t) & = & m^2_0 -0.23\cos 2\beta M^2_Z + 4x_1(t)m^2_1 
\phantom{\frac{1}{2}}\\ 
m^2_{\tilde{u}_L}(t) & = & m^2_0 + 0.35\cos 2\beta M^2_Z  
+\frac{1}{9}x_1(t)m^2_1 
+x_2(t)m^2_2 + x_3(t)m^2_3 \\
m^2_{\tilde{d}_L}(t) & = & m^2_0 -0.42\cos 2\beta M^2_Z 
+\frac{1}{9}x_1(t)m^2_1 
+x_2(t)m^2_2 + x_3(t)m^2_3 \\
m^2_{\tilde{u}_R}(t) & = & m^2_0 +0.15\cos 2\beta M^2_Z  
+\frac{16}{9} x_1(t)m^2_1 
+ x_3(t)m^2_3 \\
m^2_{\tilde{d}_R}(t) & = & m^2_0 -0.08\cos 2\beta M^2_Z 
+ \frac{4}{9}x_1(t)m^2_1 + x_3(t)m^2_3 \;\;\; .
\end{eqnarray*} Here $t=\ln \frac {M^2_X}{M^2_Z}$ and the $x_i(t)$ 
are functions that depend on the 
running of the standard model coupling constants. 
Taking  $M_X= 2 \times 10^{16}$ GeV and 
$\alpha_{GUT}=\frac{1}{24.3}$ \cite{BEOK} one has, approximately,
\[ x_1(t)=0.038;~~~x_2(t)=0.49;~~~x_3(t)=6.30. \]

For the third generation squarks, except $\tilde{b}_R$, there are additional 
contributions due to the top Yukawa coupling. 
One finds: 

\begin{eqnarray*}
m^2_{\tilde{b}_R}(t) & = & m^2_{\tilde{d}_R}(t) \\
m^2_{\tilde{b}_L}(t) & = & m^2_{\tilde{d}_L}(t) + h_0(t)m^2_0  
+ h_{ij}(t) m_i m_j \\
m^2_{\tilde{t}_R}(t) & = & m^2_{\tilde{u}_R}(t) +2h_0(t)m^2_0 
+ 2h_{ij}(t) m_i m_j+m^2_t \\
m^2_{\tilde{t}_L}(t) & = & m^2_{\tilde{u}_L}(t) +h_0(t)m^2_0 
+h_{ij}(t) m_i m_j +m^2_t \;\;\; .
\end{eqnarray*}
Here the functions $h_0(t)$ and $h_{ij}(t)$ depend on the evolution 
of the top Yukawa coupling. For example, using as input 
\footnote{It is not necessary to assume a value for $\tan \beta$ if one 
assumes \cite{COPW} that the top Yukawa coupling at the top mass is that 
which corresponds to the IR fixed point. Although the value for 
$\tan \beta$ we shall use gives a $Y_t(m_t)$ close to the fixed point 
value, we shall take $\tan \beta$ as a free parameter, and fix
the top Yukawa coupling using the known 
top mass $180\pm12$ GeV \cite{topmass}.}
$\tan \beta =2.4$ and $m_t \simeq 180 $ GeV, one has 
\begin{eqnarray*} 
h_0(t) & = & -0.494 \\ \\
h(t) & = & \left( \begin{array}{ccc}
-0.0154 & -0.0004 & -0.0024 \\
-0.0004 & -0.1214 & -0.0203 \\
-0.0024 & -0.0203 & -1.0793
\end{array} \right) \;\;\; .
\end{eqnarray*}

Because $h_0(t)$ is negative, the scalar mass contribution for $\tilde{b}_L$,
$\tilde{t}_L$ and $\tilde{t}_R$ is smaller than that for the other squarks. 
Furthermore, because $ m^2_{\tilde{d}_L}$ is very weakly dependent 
on $m^2_1$ and $h_{11}<0$, one sees that large values of the 
$U(1)$-gaugino mass will further reduce the $\tilde{b}_L$ mass 
provided that the contributions of the other two gaugino 
masses are relatively contained. This amounts to a considerably
fine--tuned cancelation of the $U(1)$ gaugino contribution
against the contributions of the other gauginos.
A large value for $m_1$ 
has the desired property of raising the neutralino masses, 
securing the role of  $\tilde{b}_L$  as the LSP. 
Taking, for example, $m_0= 20$ GeV and
\[
m_1=1900~{\rm GeV};~~~m_2=160~{\rm GeV};~~~m_3=80~{\rm GeV}
\] 
gives $m_{\tilde{b}_L}\simeq 56$ GeV, which is more than 50~GeV below the 
next  two lightest squarks ($m_{\tilde{t}_1}\simeq 115$ GeV 
and $m_{\tilde{u}_L} \simeq 260$ GeV).
\footnote{ Here ${\tilde{t}_1}$ is the lighter of the two stops and 
the detailed value  of its mass depends on the mixing between 
${\tilde{t}_L}$ and ${\tilde{t}_R}$.
This in turns depends on the value of the trilinear SUSY-breaking 
parameter $A_0$ at the GUT scale and on the supersymmetric mass 
parameter $\mu$. We have taken $A_0=0$ and have fixed $\mu$ 
from the minimization condition which must be satisfied to obtain 
$SU(2) \times U(1)$ breaking \cite{COPW}. 
For this example one has $\mu=302$ GeV.}

Once all the SUSY-breaking parameters $ m_i$ are fixed (along with the 
SUSY-preserving mass parameter $\mu$, which is then a function 
of $\tan \beta$), it is straightforward to deduce the gluino 
mass and the neutralino and chargino mass spectra. For the parameters 
detailed above, one finds:
\[
m_{\tilde{g}}= 230 ~{\rm GeV},
\]
while
\[
m_{\chi^0_1}=106~{\rm GeV};~~~m_{\chi^0_2}=305~{\rm GeV};
~~~m_{\chi^0_3}=327~{\rm GeV};~~~m_{\chi^0_4}=788~{\rm GeV}
\]
and
\[
m_{\chi^{\pm}_1}=106~{\rm GeV};~~~m_{\chi^{\pm}_2}=332~{\rm GeV}.
\]
For these same parameters the {\it tree level} Higgs masses
are 
\[
m_{H_1} = 64~{\rm GeV};~~~m_{H_2} = 532~{\rm GeV};~~~m_A = 528~{\rm GeV};~~~
 m_{H^+} = 534~{\rm GeV}.
\]
To get an idea of how ``fine tuned'' the above mass spectrum is, 
we present in  Fig.1 a scatter plot of minimal neutralino  
and chargino masses obtained by varying the input parameters 
$m_i$ and $\tan \beta$, but requiring still that 
$m_{\tilde{b}_L}\simeq 55$ GeV.  

\begin{figure}
\begin{center}
~\epsfig{file=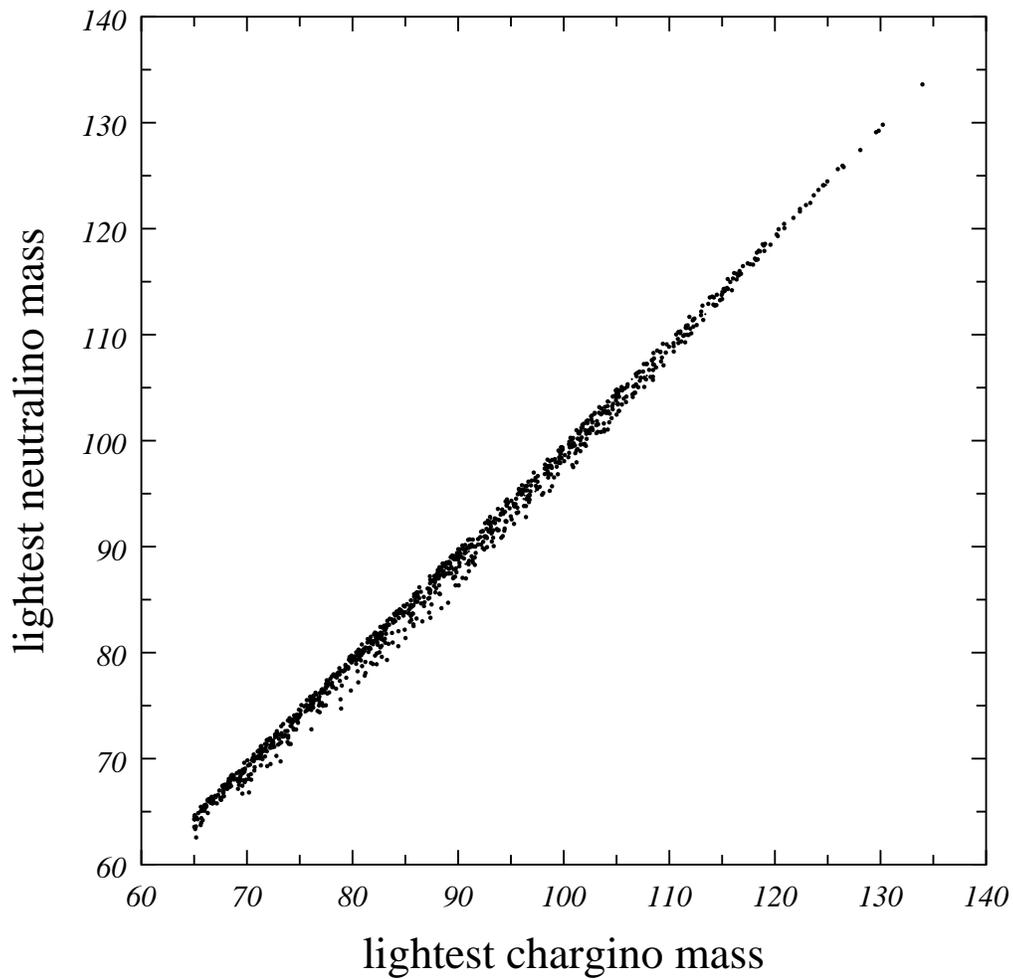,width=6in}
\end{center}
\caption{ Lightest chargino and neutralino masses obtained
for a sample of 1000 spectra with $m_{\sbotL} = 55\pm 5$ GeV. }
\end{figure}

By lifting the assumption of universal gaugino masses at the GUT scale,
we have succeeded in producing a spectrum of SUSY particles
which exhibits the essential features necessary to
account for the ALEPH ``signal'':  the LSP is a bottom squark
of mass $\sim 55$ GeV which decays to pairs of jets via $R$-parity violating 
couplings.  However, to establish the phenomenological 
viability of this scenario,
we must still examine various indirect constraints on the model.
Two such constraints are provided by the $\rho$-parameter 
\cite{Veltman} and SUSY--mediated flavor changing neutral currents.
A further set of constraints can be obtained from analyzing 
new top quark decay modes that are present in this scenario.

The supersymmetric contributions to the $\rho$ parameter have
been studied previously in Refs.~\cite{BFG,Drees}.  
It has been shown that the extra Higgs particles and the gauginos
of the SUSY standard model
give a negligible contribution to $\rho$.  However, non-degenerate 
$SU(2)_L$ squark doublets can give a large contribution to $\rho$,
in complete analogy with the similar result for quarks in the minimal
standard model.  In particular, for the spectrum described above,
the large splitting between the top and bottom squarks can give
a sizable contribution to $\rho$: writing 
$\rho = 1 +\delta\rho_{\rm MSM} + \delta\rho_{\rm SUSY}$, 
we have \cite{BFG,Drees}
\begin{eqnarray*}
\delta\rho_{\rm SUSY} &=& \frac{ 3 \alpha}{8 \pi M_W^2 \sin^2\theta_W}
\biggl[ \nonm\\
& &\ct \biggl( \cb f(\mstopa,\msbota) + \sb f(\mstopa,\msbotb)  \biggr)\nonm\\
&+&\st \biggl( \cb f(\mstopb,\msbota) + \sb f(\mstopb,\msbotb)  \biggr)\nonm\\
&-&\ct \st f(\mstopa,\mstopb) - \cb \sb f(\msbota,\msbotb) 
	\biggr],
\end{eqnarray*}
where $\theta_t$ is the mixing angle between $\stopL$ and $\stopa$,
$\theta_b$ is the mixing angle between $\sbotL$ and $\sbota$,
and $f(m_1,m_2)$ is given by
\[
f(m_1,m_2) = \frac{m_1^2+m_2^2}{2} 
           - \frac{m_1^2 m_2^2}{m_1^2 - m_2^2}\log{\frac{m_1^2}{m_2^2}}.
\]
For a top mass of $180\pm 12$ GeV, a fit to the data yields a liberal
\footnote{By this we mean that this is the maximum allowed value
of $\delta\rho$ when its correlation with the Peskin--Takeuchi
$S$ parameter \cite{Peskin} is taken into account, and
$S$ is allowed to vary such that the 90\% confidence level 
limit on $\delta\rho_{\rm SUSY}$ is maximized.  If we fix $S=0$, 
the bound decreases to 0.003.}
bound on $\delta\rho_{\rm SUSY}$ of 0.004.  
Evaluating $\delta\rho_{\rm SUSY}$ for the spectrum given
above, we find that the squarks give a significant, but not unacceptable,
contribution to the $\rho$ parameter:  $\delta\rho_{\rm SUSY}=0.0028$.

Supersymmetric particles with the mass spectrum considered here can also
have effects on flavor changing neutral currents in the $B$ system.
In particular, it is known that supersymmetry can 
give dangerously large enhancements of the $b\rightarrow s \gamma$ decay rate.
The branching ratio for charmless radiative $B$ decays has been measured
by CLEO~\cite{CLEO}; at the 95\% confidence level, they have 
reported
\[
1\times 10^{-4} < {\rm BR}( B \rightarrow X_s \gamma ) < 4 \times 10^{-4}.
\]
The contribution of supersymmetric particles to this decay rate
can be computed using the results of Refs.~\cite{Wise,Misiak,Buras,Ciuchini}.
There it was shown that
\[
\frac{{\rm BR}(B\rightarrow X_s \gamma )}{{\rm BR}(B\rightarrow X_c e \bar\nu)}
=\frac{|V_{ts}^* V_{tb}|^2}{|V_{cb}^2|}
\frac{ 6 \alpha_{\rm QED}}{\pi g(m_c/m_b)}|C_7(\mu)|^2
\]
where $g(m_c/m_b)$ is the phase space factor for the semileptonic decay,
and $C_7(\mu)$ is the coefficient of the flavor--changing operator
\[
O_7 = \frac{e}{4\pi^2} m_b \bar{s}_L\sigma_{\mu\nu}b_R F^{\mu\nu}
\]
evaluated at a scale $\mu\sim m_b$.  The QCD evolution of $C_7(\mu)$
has been computed in Refs.~\cite{Wise,Misiak}, and the contributions
of supersymmetric particles to the relevant Wilson coefficients
can be found in Ref.~\cite{Bertolini}.  In the leading logarithmic
approximation, we have \cite{Buras}
\[
C_7(\mu) \simeq \eta^{\frac{16}{23}} C_7(M_W) 
+ \frac{8}{3}\biggl(\eta^{\frac{14}{23}}-\eta^{\frac{16}{23}} \biggr) C_8(M_W)
  + \sum_{i=1}^{8} h_i \eta^{a_i},
\]
where $C_8$ is the coefficient of the chromo--magnetic moment operator
\[
O_8 = \frac{g_s}{4\pi^2} m_b \bar{s}_L\sigma_{\mu\nu} T^a b_R G^{\mu\nu}_a,
\]
and $\eta=\alpha_s(M_W)/\alpha_s(\mu)$.  
The coefficients $h_i$ and $a_i$ are pure numbers independent of
any model parameters, and can be found in Ref.~\cite{Buras}.

The coefficients $C_7(M_W)$ and $C_8(M_W)$ can be found in 
Ref.~\cite{Bertolini}.  Normalizing them appropriately,
the standard model contributions are given by
\[
C^{W^{\pm}}_{7}(M_W) = 
     \frac{3 x^3 - 2 x^2}{4(x-1)^4}\log x + 
     \frac{-8 x^3 - 5 x^2 + 7 x }{24(x-1)^3},
\]
and
\[
C^{W^{\pm}}_{8}(M_W) = 
     \frac{-3 x^2}{4(x-1)^4}\log x + \frac{-x^3 + 5 x^2 + 2 x}{8(x-1)^3},
\]
where $x = m_t^2 / M_W^2$.  In the present case, we anticipate that
the charged Higgs, the charginos, 
and the gluino will give the bulk of the supersymmetric
contribution.  The charged Higgs has long been known to give a 
sizable enhancement of radiative $B$ decay rates \cite{Wise}.
The charginos can further enhance or suppress the rate \cite{Bertolini}.
The presence of relatively small gluino and $\sbotL$ masses
in the spectrum given above indicate that the gluino may also give a sizable
contribution.  The contributions of the charged Higgs
and the charginos to $C_{7,8}(M_W)$ may be found, for instance,
in Ref.~\cite{Barbieri}.
For the gluino, we
have~\cite{Bertolini}
\[
C^{\tilde{g}}_{7}(M_W) = 
	\frac{\alpha_s \sin^2 \theta_W}{\alpha}
	\frac{U_{b\sbotL}U^*_{s\sbotL}}{V_{tb} V_{ts}^*}
	\frac{M_W^2}{m_{\tilde{b}_L}^2}
	\biggl[
	\frac{4z^2}{9(z-1)^4}\log z
     +  \frac{-4z^2-10z+2}{27(z-1)^3}
	\biggr],
\]
and
\[
C^{\tilde{g}}_{8}(M_W) = 
	\frac{\alpha_s \sin^2 \theta_W}{2 \alpha}
	\frac{U_{b\sbotL}U^*_{s\sbotL}}{V_{tb} V_{ts}^*}
	\frac{M_W^2}{m_{\sbotL}^2}
	\biggl[
	\frac{-9z+z^2}{3(z-1)^4}\log z
     +  \frac{-11z^2+40z+19}{18(z-1)^3}
	\biggr],
\]
where $z = m_{\tilde{g}}^2/m_{\sbotL}^2$, and $U_{q\tilde{q}'_L}$ is
the flavor mixing matrix that appears in the coupling of 
a quark to a gluino and a left--handed squark.  We have neglected
a small additional contribution that involves $\sbotL$--$\sbotR$
mixing.
Similar ``super-CKM'' matrices appear in the couplings of the charginos
to bottom quarks and top squarks.  Given the large non--degeneracy of the
squarks in the above spectrum, it would be reasonable to expect that
these ``super CKM'' matrices may deviate from the standard
model CKM matrix.   Hence in the following we will retain these CKM 
elements to make explicit the dependence on these angles.

Evaluating $C_{7,8}$ for the spectrum given above ($m_{\sbotL}= 56$ GeV
and $m_{\tilde{g}}=230$ GeV) and including the contribution 
of the charged Higgs and the charginos, we find an estimate for 
$\Gamma(b\rightarrow s \gamma)$:
\[
\frac{\Gamma(b\rightarrow s \gamma)[\rm SUSY]}
     {\Gamma(b\rightarrow s \gamma)[\rm MSM]} 
\simeq | 1 + 0.18 + 0.09\sigma - 0.49 \lambda |^2 
\]
where the four terms come from $W$, charged Higgs, gluino, and chargino
loops.  The contributions of {\it all} of the squarks have
been included and expressed in terms of third generation
mixing angles using CKM unitarity.  The quantities $\sigma$ and
$\lambda$ are given by
\[
\sigma = \frac{U_{b\sbotL}U^*_{s\sbotL}}{V_{tb} V_{ts}^*}
\]
and
\[
\lambda = \frac{\tilde{V}_{\tilde{t}b}\tilde{V}^*_{\tilde{t}s}}
                {V_{tb} V_{ts}^*}.
\]
The matrix $\tilde{V}$ is the super CKM matrix describing couplings
of charginos to up--type squarks and down--type quarks.
We see that for $\lambda\sim 1$ (a reasonable value),
the charginos interfere destructively with the
$W$ and charged Higgs, reducing the rate.  The possibility of such
a phenomenon has been noted in Ref.~\cite{Barbieri}.
Given the standard model estimate \cite{Ciuchini}
\[
{\rm BR}(B\rightarrow X_s\gamma)[\rm MSM]\simeq 1.9\pm 0.5\times 10^{-4},
\]
we conclude that the rate is compatible
with the CLEO determination.

The presence of light supersymmetric particles in the spectrum allows the top 
quark to decay in other modes besides the standard $t \rightarrow Wb$ mode. For
the model spectrum discussed above, the dominant non standard
decay is $t \rightarrow \tilde{b}_L \chi_1^+$ 
\footnote{ Another allowed decay is
$t\rightarrow \tilde{t}_1 \sbotL b$, but this decay is strongly kinematically
suppressed.  There may also be R-parity violating top decay modes into $bd$ 
and $bs$, but the relevant branching fractions  for these modes depend 
on the unknown couplings $\lambda_{bit}$.}. 
The branching ratio
for this mode is of O(30\%) which, although sizable, is probably 
acceptable given the experimental uncertainty in the top cross 
section \cite{topmass}. Because the dominant chargino decay is 
$\chi_1^+ \rightarrow \tilde{b}^*_L c$,  with the $\tilde{b}_L$ 
decaying into dijets, the final state for these non standard top decays 
will contain five jets. This is not a particularly easy signal to dig out. 
Nevertheless, if some top tagging can eventually be implemented at the 
Tevatron, looking for multijet decays of the accompanying $\bar{t}$ 
may be the best way to dig out the $\tilde{b}_L$ in hadronic 
interactions \footnote{Although $\tilde{b}_L-\tilde{b}_L$ pairs 
are copiously produced at the Tevatron,  because of the large QCD 
background it appears very difficult to detect them with an 
ALEPH-like analysis.}.

It is quite likely that, in the end, the ALEPH dijet enhancement will 
prove to be just a statistical fluctuation, making the scenario discussed 
here moot.  Even if this were to  turn out to be the case, we believe 
that some of our disquisitions may continue to prove useful. 
Three lessons which emerge from our speculations stand out in particular. 
First, significant deviations from the spectrum predicted by minimal 
versions \cite{BEOK,COPW} of softly broken supersymmetry can occur 
as a result of some simple changes in the underlying assumptions 
(e.g.  having non-universal gaugino masses). Second, a low mass sbottom 
(or a low mass stop) should not be unexpected, given the large top 
Yukawa coupling. Third, although there is a natural prejudice against 
R-parity violating couplings, their presence at some level is by no 
means ruled out. If present, such couplings totally alter the 
``standard'' signals of supersymmetry.
\vskip 1 cm
\noindent{\Large{\bf{Acknowledgments}}}

This work was supported in part by the Department of Energy under Grant No
FG03-91ER40662, Task C.

\pagebreak

\end{document}